\def\tr{{\text{tr}}\,}

\def\be{\begin{equation}}
\def\ee{\end{equation}}
\def\bea{\begin{eqnarray}}
\def\eea{\end{eqnarray}}
\def\bse{\begin{subequations}}
\def\ese{\end{subequations}}

\documentclass[prb,twocolumn,showpacs,amsmath,amssymb,eqsecnum]{revtex4}
\usepackage{graphicx}% Include figure files
\usepackage{dcolumn}% Align table columns on decimal point
\usepackage{bm}% bold math
\begin{document}
\title{Quantum critical behavior in disordered itinerant ferromagnets:\\
	Instability of the ferromagnetic phase
       %\small{$[$ Phys. Rev. B {\bm 63}, 174428 (2001) $]$
}
\author{Sharon L. Sessions}
\affiliation{Max-Planck-Institute for Physics of Complex Systems,
         D-01187 Dresden, Germany}
\author{D.Belitz}
\affiliation{Department of Physics and Materials Science Institute, %\\
         University of Oregon, %\\
         Eugene, OR 97403}
\date{\today}

\begin{abstract}
An effective field theory is derived that describes the quantum critical
behavior of itinerant ferromagnets as the transition is approached from
the ferromagnetic phase. This
complements a recent study of the critical behavior on the paramagnetic
side of the phase transition,
and investigates the role of the ferromagnetic Goldstone modes near
criticality. We find that the Goldstone
modes have no direct impact on the critical behavior, and that the
critical exponents are
the same as determined by combining results from the paramagnetic phase
with scaling arguments.
\end{abstract}


\maketitle

\section{Introduction}
\label{sec:I}

In two recent papers,\cite{us_fm_local_I,us_fm_local_II} hereafter denoted
by I and
II, respectively, an effective field theory has been developed that
describes the quantum ferromagnetic transition in disordered itinerant
electron systems, and the exact critical behavior was determined for
all spatial dimensions $d>2$. Soft modes in addition to the
order parameter fluctuations, viz., diffusive particle-hole excitations
(``diffusons''), were found to play a crucial role in the
determination of the critical behavior. This theory studied
the transition as it is approached from the paramagnetic phase. The
critical behavior in the ferromagnetic phase, in particular the exponent
$\beta$, was obtained from scaling arguments. This raises the following
question: In the ferromagnetic phase of a Heisenberg magnet, the existence
of spin waves, which are the Goldstone modes that result from the spontaneous
breaking of the rotational symmetry in spin space, changes the soft-mode
spectrum compared to the paramagnetic phase. Given the importance of the
soft modes for the critical behavior, will the Goldstone modes invalidate
the scaling arguments based on a theory for the paramagnetic phase?

A related question follows from the fact that
the Goldstone modes, contrary to the soft particle-hole excitations,
exist at nonzero temperature as well as at $T=0$.
Therefore, if there were an observable whose leading
critical behavior depended on the Goldstone modes, then one would expect
that observable to exhibit classical critical behavior on the
ferromagnetic side of the phase transition, even at temperatures
low enough for the system to otherwise exhibit quantum critical behavior.
In such a case, the scaling arguments used in II might still be formally
correct, but their region of validity would vanish.

In the present paper we investigate the validity of the scaling arguments
used before. We will find that they were correct,
and the critical behavior obtained in II was indeed
exact, even on the ferromagnetic side of the transition. To this end, we
develop the ferromagnetic analog of the theory described in I and II.
The general strategy is to derive a theory that expresses the system in
terms of the order parameter fluctuations, and any other soft modes that
couple to them. We then analyze the resulting
theory to determine the critical behavior of the system as it approaches
the transition from the ferromagnetic phase.

This paper is organized as follows. In Sec.\ \ref{sec:II} we
present a simple generalized mean-field theory that allows an explicit
description of the ferromagnetic phase based on the theory developed
in I. Its results agree with those obtained in II, except that it does
not reproduce the logarithmic corrections to power-law scaling found
in the latter paper. In Sec.\ \ref{sec:III} we present and motivate
an effective action for itinerant quantum ferromagnets that describes
the instability of the ferromagnetic phase. This action is analyzed
by means of renormalization-group methods in Sec.\ \ref{sec:IV}. It
is shown that the generalized mean-field theory yields indeed the
exact critical behavior except for logarithms. In Sec.\ \ref{sec:V}
we summarize and discuss our results. Several technical
points related to a derivation of the effective action from a
microscopic model are relegated to an appendix.

\section{Generalized mean-field theory}
\label{sec:II}

In this section we present a simple way to obtain an approximate theory
for the ferromagnetic phase from the formalism of I. It will turn out
that this simple theory produces essentially the correct result, and
the technically more involved development in the later sections
will serve to show this.

\subsection{A simplified effective action}
\label{subsec:II.A}

We recall that the effective action ${\cal A}_{\rm eff}$ in I took the
form of a
Landau-Ginzburg-Wilson (LGW) functional of the order parameter field
${\bm M}$, a generalized nonlinear sigma-model for the soft fermionic
degrees of freedom $q$, and a term that couples these fields,
\be
{\cal A}_{\rm eff} = {\cal A}_{\rm LGW}[{\bm M}]
                     + {\cal A}_{{\rm NL}\sigma{\rm M}}[q]
                     + {\cal A}_{\rm c}[{\bm M},q]\quad.
\label{eq:2.1}
\ee
We simplify the expressions given in I for these terms by making a
mean-field approximation for the order parameter field, i.e., we
replace the fluctuating vector field ${\bm M}$ by its expectation
value, which is characterized by a number $m$,
\bea
{\bm M}^{\alpha}({\bm x},\tau) &\approx&
   \langle {\bm M}^{\alpha}({\bm x},\tau) \rangle = m\,{\hat z}\quad,
\nonumber\\
{^iM}_n^{\alpha}({\bm k}) &\approx&
        \delta_{i3}\,\delta_{n0}\,\sqrt{V/T}\,m\quad.
\label{eq:2.2}
\eea
Here $({\bm x},\tau)$ denote the dependence of ${\bm M}$ on real space
and imaginary time, ${\bm k}$ and $n$ are the wavevector and the
Matsubara frequency index, respectively, of the Fourier transform of
${\bm M}$, $\alpha$ is a replica index, $V$ denotes the system volume,
and we have assumed that the magnetization orders in the $z$ or $3$-direction.
With this approximation, the LGW part of the action becomes simply a
Landau theory,
\be
{\cal A}_{\rm LGW} \approx \frac{V}{T}\left[\frac{t}{2}\,m^2
        + \frac{u}{4}\,m^4 + O(m^6)\right] \quad.
\label{eq:2.3}
\ee
The coupling term in the action (Eq.\ (2.25c) in I) simplifies to
\bea
{\cal A}_{\rm c} &\approx& m\,\sqrt{\pi K_t} \int d{\bm x}\sum_{\alpha}
                           \sum_{r=0,3} (\sqrt{-1})^r
\nonumber\\
&&\times \sum_m \tr\left[(\tau_r\otimes s_3)\,{\hat Q}^{\alpha,\alpha}_{mm}
   ({\bm x})\right] \quad.
\label{eq:2.4}
\eea
Here $K_t$ is the spin-triplet interaction amplitude of the underlying
fermionic model; the spin-triplet interaction term has been decoupled
by introducing the Hubbard-Stratonovich field ${\bm M}$. ${\hat Q}$ is
a composite field\cite{us_fermions_I} that originates from the fermion
variables,
\footnote{We use the notation, $a\approx b$ for ``$a$ approximately
 equals $b$'', $a\protect\cong b$ for ``$a$ is isomorphic to $b$'', and
 $a \sim b$ for ``$a$ scales like $b$''.}
\bse
\label{eqs:2.5}
\bea
Q_{12} \cong \frac{i}{2}\hskip -1pt\left(\hskip -3pt \begin{array}{cccc}
           -\psi_{1\uparrow}{\bar\psi}_{2\uparrow} &
               -\psi_{1\uparrow}{\bar\psi}_{2\downarrow} &
                  -\psi_{1\uparrow}\psi_{2\downarrow} &
                        \ \ \psi_{1\uparrow}\psi_{2\uparrow}  \\
          -\psi_{1\downarrow}{\bar\psi}_{2\uparrow} &
             -\psi_{1\downarrow}{\bar\psi}_{2\downarrow} &
                 -\psi_{1\downarrow}\psi_{2\downarrow} &
                      \ \ \psi_{1\downarrow}\psi_{2\uparrow}  \\
          \ \ {\bar\psi}_{1\downarrow}{\bar\psi}_{2\uparrow} &
             \ \ {\bar\psi}_{1\downarrow}{\bar\psi}_{2\downarrow} &
                 \ \ {\bar\psi}_{1\downarrow}\psi_{2\downarrow} &
                      -{\bar\psi}_{1\downarrow}\psi_{2\uparrow} \\
          -{\bar\psi}_{1\uparrow}{\bar\psi}_{2\uparrow} &
             -{\bar\psi}_{1\uparrow}{\bar\psi}_{2\downarrow} &
                 -{\bar\psi}_{1\uparrow}\psi_{2\downarrow} &
                      \ \ {\bar\psi}_{1\uparrow}\psi_{2\uparrow} \\
                    \end{array}\hskip -2pt \right)\ .
\nonumber\\
\label{eq:2.5a}
\eea
Here the $\psi$ and ${\bar\psi}$ are the fermionic, i.e., Grassmann-valued
fields that provide the basic description of the
electrons,\cite{Negele_Orland_88}
and all fields are understood to be taken at position ${\bm x}$. The indices
$1$, $2$, etc. denote the dependence of the Grassmann fields on fermionic
Matsubara frequencies $\omega_{n_1} = 2\pi T(n_1+1/2)$ and replica indices
$\alpha_1$, etc., and the arrows denote the spin projection. It is convenient
to expand the $4\times 4$ matrix in Eq.\ (\ref{eq:2.5a}) in a
spin-quaternion basis,\cite{Efetov_Larkin_Khmelnitskii_80}
\be
Q_{12} = \sum_{r,i=0}^{3}(\tau_r\otimes s_i)\,{_r^iQ}_{12}\quad,
\label{eq:2.5b}
\ee
with $\tau _0 = s_0 = \openone_2$ the $2\times 2$ unit matrix, and
$\tau_j = -s_j = -i\sigma_j$, $(j=1,2,3)$, with $\sigma_{1,2,3}$ the Pauli
matrices. In this basis, $i=0$ and $i=1,2,3$ describe the spin-singlet and
spin-triplet degrees of freedom, respectively. The $r=0,3$ components
correspond to the particle-hole channel (i.e., products of ${\bar\psi}\psi$),
while $r=1,2$ describe the particle-particle channel (i.e., products
of ${\bar\psi}{\bar\psi}$ or $\psi\psi$). For our purposes the
latter are not important, and we therefore drop the $r=1,2$ from the
spin-quaternion basis, as was done in I and II. In terms of the remaining fields,
the spin density can be expressed as
\bea
n_{\rm s}^i({\bm x},i\Omega_n) &=& \sqrt{T}\sum_m\sum_{ab}
   {\bar\psi}_{m,a}({\bm x})\,\sigma^i_{ab}\,\psi_{m+n,b}({\bm x})
\nonumber\\
&&\hskip -40pt = \sqrt{T}\sum_m\sum_{r=0,3}(\sqrt{-1})^r \tr\,
   \left[(\tau_r\otimes s_i)\,Q_{m,m+n}({\bm x})\right]\quad,
\nonumber\\
\label{eq:2.5c}
\eea
and the magnetization couples linearly to the spin density, cf.
Eq.\ (\ref{eq:2.4}). The difference between $Q$ and ${\hat Q}$ is
that the former contains all fermionic degrees of freedom, while
the latter contains the soft fermion excitations only. The latter
can be parametrized in terms of a matrix field
$q$,\cite{us_fermions_I,us_fm_local_I}
\be
{\hat Q} = \left( \begin{array}{cc}
                 \sqrt{1-qq^{\dagger}} & q   \\
                    q^{\dagger}        & -\sqrt{1-q^{\dagger} q} \\
           \end{array} \right)\quad.
\label{eq:2.5d}
\ee
\ese%
The elements of this block matrix, clockwise from upper left, denote those
elements of ${\hat Q}$ that have frequency labels $n_1,n_2>0$; $n_1>0,n_2<0$;
$n_1,n_2<0$; and $n_1<0,n_2>0$, respectively.

The nonlinear sigma-model can be expressed in terms of ${\hat Q}$
as\cite{Wegner_79,Efetov_Larkin_Khmelnitskii_80,Finkelstein_83,us_fermions_I,
us_fm_local_I}
\bse
\label{eqs:2.6}
\bea
{\cal A}_{{\rm NL}\sigma{\rm M}}&=&\frac{-1}{2G}\int d{\bm x}\
     \tr\left(\nabla\hat Q ({\bm x})\right)^2
\nonumber\\
&&\hskip -50pt + 2H \int d{\bm x}\ \tr\left(\Omega\,{\hat Q}({\bm x})
     \right)
 + {\cal A}_{\rm int}^{(s)}[{\hat Q}]\quad,
\label{eq:2.6a}
\eea
with $\Omega$ a frequency matrix with elements
\be
\Omega_{12} = \left(\tau_0\otimes s_0\right)\,\delta_{12}\,\omega_{n_1}\quad.
\label{eq:2.6b}
\ee
The coupling constants $G$ and $H$ are proportional to the inverse
conductivity and the specific heat coefficient, respectively.
${\cal A}_{\rm int}^{(s)}$ is the spin-singlet interaction,
\bea
{\cal A}_{\rm int}^{\,(s)}&=&\frac{\pi TK_{\rm s}}{4}\int d{\bm x}
   \sum_{r=0,3}(-1)^r \sum_{n_1,n_2,m}\sum_\alpha
\nonumber\\
&&\times\left[\tr \left((\tau_r\otimes s_0)\,\hat{Q}_{n_1,n_1+m}^{\alpha\alpha}
({\bm x})\right)\right]
\nonumber\\
&&\times\left[\tr \left((\tau_r\otimes s_0)\,\hat{Q}_{n_2+m,n_2}^{\alpha\alpha}
({\bm x})\right)\right]\quad,
\label{eq:2.6c}
\eea
with a coupling constant $K_{\rm s}$.
At the level of the bare effective action, ${\cal A}_{{\rm NL}\sigma{\rm M}}$
does not contain any spin-triplet interaction term, since the 
Hubbard-Stratonovich decoupling turned the latter into the terms containing 
$\bm M$. For later reference we note, however, that under renormalization a 
spin-triplet interaction will be generated as long as $K_s\neq 0$. We therefore 
add to the action a term
\bea
{\cal A}_{\rm int}^{\,(t)}&=&\frac{\pi T{\tilde K}_t}{4}\int d{\bm x}
    \sum_{r=0,3} (-1)^r \sum_{n_1,n_2,m}\sum_\alpha \sum_{i=1}^3
\nonumber\\
&&\times\left[\tr \left((\tau_r\otimes s_i)\,\hat{Q}_{n_1,n_1+m}^{\alpha\alpha}
({\bm x})\right)\right]
\nonumber\\
&&\times\left[\tr \left((\tau_r\otimes s_i)\,\hat{Q}_{n_2+m,n_2}^{\alpha\alpha}
({\bm x})\right)\right]\quad,
\label{eq:2.6d}
\eea
\ese%
with ${\tilde K}_t$ the spin-triplet interaction constant generated by
renormalization.

\subsection{Generalized Landau theory}
\label{subsec:II.B}

We now can obtain a generalized mean-field or Landau free energy density,
$f(m)$, that takes into account the effects of the fermionic soft modes by
formally integrating out $q$,
\be
f(m) = \lim_{N\rightarrow 0} \frac{-T}{VN}\,\ln\int D[q]\ e^{{\cal A}_{\rm eff}}
       \quad,
\label{eq:2.7}
\ee
with $N$ the number of replicas. Since ${\cal A}_{\rm eff}$ contains $q$ to
all orders, the integral in Eq.\ (\ref{eq:2.7}) can be performed only in
terms of a loop expansion. The lowest-order term is obtained by expanding
${\hat Q}$ to second order in $q$. The integration can then be done
exactly. The Gaussian propagators that are needed for this calculation
have been given in I (the generalization to the case ${\tilde K}_t\neq 0$ is
straightforward, and can also be found in Ref.\ \onlinecite{us_R}),
and we therefore only give the result,
\begin{widetext}
\be
f(m) \approx f(m=0) + \frac{t}{2}\,m^2 + \frac{u}{4}\,m^4
   + \frac{2}{V}\sum_{\bm k} T\sum_{n=1}^{\infty} \ln \left(
   \frac{\left({\bm k}^2 + G(H+{\tilde K}_t)\Omega_n\right)^2 + \pi G^2K_t m^2}
        {({\bm k}^2 + GH\Omega_n)^2 + \pi G^2 K_t m^2}\right)\quad.
\label{eq:2.8}
\ee
\end{widetext}
By minimizing $f(m)$ with respect to $m$, we obtain a generalized mean-field
equation of state. If we introduce suitable units, drop non-essential
constants, and add an external magnetic field $h$, the latter can be written
\bse
\label{eqs:2.9}
\bea
h &=& t\,m + u\,m^3
\nonumber\\
&&\hskip 17pt - m\,\frac{\rm const}{V}\sum_{\bm k}
    T\sum_{n=1}^{\infty} \frac{({\bm k}^2 + \Omega_n)\Omega_n}
                              {\left[({\bm k}^2 + \Omega_n)^2 + m^2\right]^2}
                                                \quad,
\nonumber\\
\label{eq:2.9a}
\eea
with ${\rm const}>0$.
An inspection of the integral shows that the leading term for small $m$
is finite for $d>2$, and that the leading nonanalytic $m$-dependence
is given by $m^{(d-2)/2}$. The generalized mean-field equation of state
thus reads
\be
h = t\,m + v\,m^{d/2} + u\,m^3\quad,
\label{eq:2.9b}
\ee
\ese%
with $v>0$.

We see that for $2<d<6$ the nonanalytic term dominates over the $m^3$
contribution. The exponents $\beta$ and $\delta$, defined by
$m(h=0) \propto (-t)^{\beta}$ and $m(t=0) \propto h^{1/\delta}$, respectively,
for $2<d<6$ in this approximation are
\be
\beta = 2/(d-2)\quad,\quad \delta = d/2\quad.
\label{eq:2.10}
\ee
These values agree with the ones found in II, apart from logarithmic
corrections to scaling that the generalized mean-field theory misses.
Equation (\ref{eq:2.9a}) is also very similar to the effective equation
of state that was obtained in Ref.\ \onlinecite{us_fm_dirty}, and it has
the same qualitative properties. Notice, however, that the derivation
of Eq.\ (\ref{eq:2.9a}) did not involve any divergent integrals, while
the equivalent result in Ref.\ \onlinecite{us_fm_dirty} was obtained by
resumming an infinite series of divergent terms. We will discuss the
relation between these theories in more detail in Sec.\ \ref{sec:V}.

While the above procedure provides a fast and simple way to incorporate
soft-mode effects into the equation of state, and the agreement with
prior results is encouraging, our mean-field approximation is of course
uncontrolled. Furthermore, since it neglects the order parameter
fluctuations, it does not shed any light on the role played by the
Goldstone modes. In the following two sections we will therefore
construct an effective field theory for the ferromagnetic phase, and
perform a renormalization-group analysis to study the critical behavior.
We will find that the results of the generalized mean-field theory are
indeed exact apart from logarithmic corrections to scaling.

\section{Effective field theory for ferromagnets}
\label{sec:III}

In this section we present an effective field theory for the
ferromagnetic phase in analogy to the one for the paramagnetic
phase given in I. The structure of
such a theory can be deduced without a calculation by combining
various existing results for disordered itinerant ferromagnets.
In the interest of brevity and clarity, we pursue this route.
We have also checked the result by means of a derivation from a
microscopic model, using the methods of I and
Ref.\ \onlinecite{us_fm_mit_I}. We summarize the salient points
of this derivation in Appendix\ \ref{app:A}; a
complete account has been given in Ref.\ \onlinecite{sls_thesis}.

In I it was shown how to derive an effective long-wavelength and
low-frequency field theory for a disordered metal that approaches
a ferromagnetic instability. We now give the corresponding action
for a ferromagnetic metal, which we will motivate below. We denote
the fluctuations of the longitudinal part of the order parameter
field ${\bm M}$ about its expectation value $m$ by $\delta M_{\ell}$,
and the transverse part by $M_t$. Similarly, we denote the spin-singlet
and the longitudinal spin-triplet components of the fermionic field
$q$ by $q_{\ell}$, and the transverse spin-triplet components by $q_t$.
The action then has the general form
\bse
\label{eqs:3.1}
\bea
\label{eq:3.1a}
{\cal A}_{\rm eff}[{\bm M},q] &=& 
             {\cal A}^{\rm G}_{\ell}[\delta M_{\ell},q_{\ell}]
    + {\cal A}^{\rm G}_t[M_t,q_t]
\nonumber\\
&& + \Delta{\cal A}[\delta M_{\ell},M_t,q_{\ell},q_t] \quad.
\eea
Here the superscript G denotes the action at the Gaussian level, where
the longitudinal and transverse fields decouple,
separated into longitudinal and transverse contributions.
$\Delta{\cal A}$ represents contributions of higher than bilinear order in
the fields. We now specify the various terms in this action
in the same schematic notation is in I (cf. Eq.\ (3.8) of I), suppressing
everything that is not necessary for power counting, and considering all
fields as functions of real space position $\bm x$ and Matsubara frequencies.
The Gaussian contributions are then given by
\bea
{\cal A}_{\ell}^{\rm G} &=& \int d{\bm x}\ \left[
      \vert t\vert + a_{\ell,d-2}\,\partial_{\bm x}^{d-2}
	  + a_{\ell,2}\,\partial_{\bm x}^2 \right]\,(\delta M_{\ell})^2
\nonumber\\
&&+ \int d{\bm x}\ \left[ \frac{1}{ G_{\ell}}\,\partial_{\bm x}^2
	+ H_{\ell}\,\Omega + TK\right]\ q_{\ell}^2
	\nonumber\\
&&+ \ c_{\ell,1}\sqrt{T} \int d{\bm x}\ \delta M_{\ell}\,q_{\ell}
	\quad,
\label{eq:3.1b}\\
{\cal A}_{t}^{\rm G} &=& \int d{\bm x}\ \left[
	a_{t,d-2}\,m^{(d-4)/2}\partial_{\bm x}^2
      + a_{t,2}\partial_{\bm x}^2 \right]\ M_t^2
	\nonumber\\
&&+ \int d{\bm x}\  \left[ \frac{1}{ G_t}\,\partial_{\bm x}^2
	+H_t\Omega + K m \right]\ q_t^2
	\nonumber\\
&&+\ c_{t,1}\sqrt{T}\int d{\bm x}\ M_t\,q_t
	\quad.
\label{eq:3.1c}
\eea
\ese%
Here $K$ is a generic coupling constant that is equal to $K_{\rm s}$ in the 
spin-singlet channel, ${\tilde K}_{\rm t}$ in the longitudinal spin-triplet 
channel, and $K_{\rm t}$ in the transverse spin-triplet channels.
We now motivate this form of the Gaussian effective action.

First of all, the magnetic order breaks the rotational symmetry in spin
space and forces us to distinguish between the longitudinal and transverse
components of all fields. The soft-mode structures in the longitudinal
channels are similar to those in a paramagnet. The spin-singlet and the
longitudinal spin-triplet fermionic fields are known to be unaffected by
the presence of a nonzero magnetization (see, e.g., Ref.\ \onlinecite{us_R});
the vertex for $q_{\ell}$ thus has the same form as in I.
The same is true for the longitudinal order parameter fluctuations. The
distance from criticality, $t$, is actually magnetization dependent, but this
is irrelevant for power counting purposes. The coupling between
$\delta M_{\ell}$ and $q_{\ell}$ is also unaffected. Notice that this coupling
provides the leading frequency dependence of the $\delta M_{\ell}$ propagator,
viz., a term proportional to $\Omega/{\bm k}^2$ with $\bm k$ a wavevector.
We will refer to this propagator as the longitudinal magnon.
That is, in Fourier space and for small wavenumbers and frequencies
the longitudinal magnon propagator has the form
\bea
\langle \delta M_{\ell}({\bm k})\,\delta M_{\ell}(-{\bm k})\rangle &=&
\nonumber\\
&&\hskip -99pt \frac{1}
   {\vert t\vert + a_{\ell,d-2}\vert{\bm k}\vert^{d-2}
      + a_{\ell,2}\,{\bm k}^2 + G_{\ell}\,c_{\ell,1}^2\,\Omega/{\bm k}^2}\quad.
\label{eq:3.2}
\eea
For this reason we have left the less leading direct frequency dependence
($\propto\Omega$) out of the $\delta M_{\ell}$ vertex.

The Eqs. (\ref{eqs:3.1}) and (\ref{eq:3.2}) demonstrate an important point that
was discussed in detail in I and II. Namely, there is more than one time scale
in the problem, and hence more than one dynamical exponent $z$.
\footnote{$z$ is defined by the scaling relation between frequency and
wavenumber, $\Omega \sim \vert{\bm k}\vert^z$.}
The diffusive dynamics of the fermions are described by an exponent $z=2$,
while the critical $z$, which describes the critical dynamics of the order
parameter field, has a different value. This will be important in what follows.

The soft-mode structure in the transverse spin-triplet channel is
qualitatively different compared to the paramagnetic phase. The transverse
fermion fields acquire a mass proportional to $m$, see, Ref.\ \onlinecite{us_R}
or Eq.\ (\ref{eq:2.4}).
The order parameter vertex, on the other hand, is massless due to the
presence of ferromagnetic Goldstone modes. The structure of the fermionic
vertex suggests that the magnetization can scale like a gradient squared.
For the transverse magnetization vertex, we expect terms that
represent the classic dispersion relation,\cite{Moriya_85}
\bse
\label{eqs:3.3}
\be
\label{eq:3.3a}
\Omega \sim m {\bm k}^2 \quad,
\ee
as well as terms that reflect the nonanalytic magnetization dependence of the
magnon effective mass,\cite{us_magnon_dispersion}
\be
\label{eq:3.3b}
\Omega \sim m^{(d-2)/2}{\bm k}^2\quad.
\ee
\ese
Again the leading dynamics come from the coupling to the fermions, which here
produces a term $\Omega/m$ due to the mass in the $q_t$ vertex.
Eqs. (\ref{eqs:3.3}) therefore show that we must
include a simple gradient-sqared term, as well as a gradient-squared term with
a coefficient proportional to $m^{(d-4)/2}$ in the $M_t$ vertex in order
to accurately represent the Goldstone modes. The latter reflects the same
nonanalyticity as the $\partial_{\bm x}^{d-2}$ term in the longitudinal
channel (recall that $m \sim \partial_{\bm x}^2$). The transverse magnon
or Goldstone mode thus has the form
\bea
\langle M_t({\bm k})\,M_t(-{\bm k})\rangle &=&
\nonumber\\
&&\hskip -80pt \frac{1}{a_{t,d-2}\,m^{(d-4)/2}\,{\bm k}^2 + a_{t,2}\,{\bm k}^2
                        + c_{t,1}^2\,\Omega/m}\quad.
\label{eq:3.4}
\eea

Beyond Gaussian order, the transverse and singlet/longitudinal spin
channels are coupled. By analyzing the spin structure of the action as given
in I, we find, schematically,
\begin{widetext}
\bea
\Delta{\cal A} &=& c_{2,1} \sqrt{T} \int d{\bm x}\ \delta M_{\ell}\,q_{\ell}^2
                  + c_{2,2} \sqrt{T} \int d{\bm x}\ \delta M_{\ell}\,q_t^2
                  + c_{2,3} \sqrt{T} \int d{\bm x}\ \delta M_t\,q_{\ell}\,q_t
\nonumber\\
&& + u_1\,T\int d{\bm x}\ (\delta M_{\ell})^4
   + u_2\,T\int d{\bm x}\ (\delta M_{\ell})^2\,(\delta M_t)^2
    + u_3\,T\int d{\bm x}\ (\delta M_t)^4
\nonumber\\
&& + \int d{\bm x}\ \left[\frac{1}{ G_{4,1}}\,\partial_{\bm x}^2
	+ H_{4,1}\,\Omega\right]\,q_{\ell}^4
   + \int d{\bm x}\ \left[\frac{1}{ G_{4,2}}\,\partial_{\bm x}^2
	+ H_{4,2}\,\Omega + m\right]\,q_{\ell}^2\, q_t^2
\nonumber\\
&&\hskip 140pt + \int d{\bm x}\ \left[\frac{1}{ G_{4,3}}\,\partial_{\bm x}^2
               + H_{4,3}\,\Omega + m\right]\,q_t^4\quad.
\label{eq:3.5}
\eea
\end{widetext}
The first and the third class of terms are the $Mq^2$ and $q^4$ vertices,
respectively, that were found to be important in describing the critical
behavior as the transition is approached from the paramagnetic phase.
The $Mq^2$-vertices would generate the nonanalyticities in the
$M^2$-vertices if they had not been included in the bare action, and
together with the $q^4$-vertices they give rise to log-log-normal
corrections to power laws. We also include an $M^4$-vertex since its
coupling constant, $u$, is a dangerously irrelevant operator with respect
to the magnetization.\cite{Ma_76}

\section{Renormalization-group analysis}
\label{sec:IV}

We now conduct a power-counting analysis of the action
given in Sec.\ \ref{sec:III}.  The purpose of this exercise is
to determine the role of the Goldstone modes in the
ferromagnetic transition, and thus compare the critical behavior on
the ferromagnetic side of the transition with the results from the
paramagnetic theory.

For the following analysis, we assign lengths $L$ a scale dimension of
$[L]=-1$. Rescaling
lengths by a factor $b$ under a renormalization-group transformation
will change all other quantities according to $A\rightarrow b^{[A]}A$,
with $[A]$ the scale dimension of $A$.  In particular, the
temperature $T$ and frequency $\Omega$ have a scale dimension
$[T]=[\Omega]\equiv z$. The scale dimensions of the fields we characterize
as usual by means of exponents $\eta$,\cite{Ma_76}
\bse
\label{eqs:4.0}
\bea
[\delta M_{\ell}] &=& (d-2+\eta_{\ell})/2\quad,
\label{eq:4.0a}\\
\left[M_t\right] &=& (d-2+\eta_t)/2\quad,
\label{eq:4.0b}\\
\left[q_{\ell,t}\right] &=& (d-2+\eta'_{\ell,t})/2\quad.
\label{eq:4.0c}
\eea
\ese
As in Sec. \ref{sec:III} we consider all fields functions of frequency and
real space position. Throughout this section we use Ma's technique
of choosing scale dimensions, and then checking self-consistently whether
this choice leads to a physical fixed point.\cite{Ma_76}

\subsection{Stable fixed point}
\label{subsec:IV.A}

Before considering the critical fixed point that describes the ferromagnetic
transition, it is illustrative to discuss the stable fixed point that 
corresponds to the system deep in the ferromagnetic phase. We start with the 
longitudinal degrees of freedom.

The natural choice for $t$ is to be marginal at the stable fixed 
point.\cite{Ma_76} This choice determines the scale dimension of 
$\delta M_{\ell}$, so we have
\bse
\label{eqs:4.1}
\be
[t] = 0\quad,\quad \eta_{\ell} = 2\quad.
\label{eq:4.1a}
\ee
$q_{\ell}$ we expect to be unaffected by the magnetization, so we choose its
scale dimension to be consistent with diffusive behavior,
\be
\eta'_{\ell} = 0\quad.
\label{eq:4.1b}
\ee
By the same argument, the dynamics of $q_{\ell}$, i.e., the scale dimensions
of $\Omega$ and $T$ in the $q_{\ell}$ vertex, should be governed by a
dynamical exponent $z=2$. We further choose the coupling $c_{\ell,1}$ to be
marginal, which implies that the dynamics of $\delta M_{\ell}$, as represented
by the factor $\sqrt{T}$ in the last term in Eq.\ (\ref{eq:3.1b}), are also
governed by a $z=2$. We thus have
\be
[c_{\ell,1}] = 0\quad,\quad z_{\rm diff} = z_{\ell} = 2\quad.
\label{eq:4.1c}
\ee
From Eq.\ (\ref{eq:3.1b}) we then find that $G_{\ell}$, $H_{\ell}$,
and $K$ are
all marginal,
\be
[G_{\ell}] = [H_{\ell}] = [K] = 0\quad,
\label{eq:4.1d}
\ee
\ese
while $a_{\ell,d-2}$ and $a_{\ell,2}$ are irrelevant with scale dimensions
$[a_{\ell,d-2}] = -(d-2)$ and $[a_{\ell,2}] = -2$, respectively.

We now turn to the transverse degrees of freedom. $q_t$ is expected to be a
massive fluctuation deep in the ferromagnetic phase, which means we choose
\bse
\label{eqs:4.2}
\be
\eta'_t = 2\quad.
\label{eq:4.2a}
\ee
This, together with the marginality of $K$, implies that $m$ is marginal,
as one would expect on physical grounds,
and so is $h$, since $[h] = [m] + [t]$, see the equation of state,
Eqs.\ (\ref{eqs:2.9}),
\be
[m] = [h] = 0\quad,
\label{eq:4.2b}
\ee
while $1/G_t$ and $H_t$ are irrelevant with scale dimensions
$[1/G_t] = [H_t] = -2$. We further choose $a_{t,2}$ and $c_{t,1}$
to be marginal. This renders $a_{t,d-2}$ marginal as well,
\be
[a_{t,2}] = [a_{t,d-2}] = [c_{t,1}] = 0\quad,
\label{eq:4.2c}
\ee
and it leads to
\be
\eta_t = 0\quad,
\label{eq:4.2d}
\ee
and to a transverse time scale governed by a dynamical exponent
\be
z_t = 2\quad.
\label{eq:4.2e}
\ee
\ese

We see that at the stable fixed point the various time scales coincide,
and there is only one dynamical exponent, $z=2$. This reflects the fact that,
deep in the ferromagnetic phase,
the diffusive electron dynamics, the frequency dependence of the longitudinal
magnon propagator, Eq.\ (\ref{eq:3.2}), and the dispersion of the Goldstone
modes, Eqs.\ (\ref{eqs:3.3}), all lead to the same scaling of
the frequency with the wavenumber. From Eq.\ (\ref{eq:3.5})
it is then easily seen that all non-Gaussian terms in the action are
irrelevant with respect to the stable fixed point, with the least irrelevant
operators having scale dimensions equal to $-(d-2)/2$. Temperature and frequency
are relevant operators, of course. This
establishes the description of the ferromagnetic phase within our formalism.

\subsection{Critical fixed point}
\label{subsec:IV.B}

We now turn to the critical fixed point that describes the
ferromagnet-to-paramagnet transition. Since this is a symmetry-restoring
transition, any sensible candidate for the critical fixed point must have the
feature that the longitudinal and transverse fields have the same scale
dimensions. We demand that the fermion fields be diffusive,
\bse
\label{eqs:4.3}
\be
\eta'_{\ell} = \eta'_t = 0\quad,
\label{eq:4.3a}
\ee
with a time scale given by
\be
z_{\rm diff} = 2\quad.
\label{eq:4.3b}
\ee
\ese
Equations (\ref{eq:3.1b},\ref{eq:3.1c}) then imply
\be
[G_{\ell}] = [G_t] = [H_{\ell}] = [H_t] = [K] = 0\quad.
\label{eq:4.4}
\ee
We furthermore require that $c_{1,\ell}$ and $c_{1,t}$ are marginal,
which implies
\be
[\delta M_{\ell}] = [M_t] = 1 + (d-z)/2\quad.
\label{eq:4.5}
\ee
Here $z$ is the dynamical exponent associated with the factor of $\sqrt{T}$
in either of the coupling terms in Eqs.\ (\ref{eq:3.1b},\ref{eq:3.1c}). As
was explained in I, this $z$ can equal to $z_{\rm diff}$, e.g. in terms
in perturbation theory where the longitudunal magnon propagator is convoluted 
with a diffusive one, which makes the frequency in Eq.\ (\ref{eq:3.2}) scale
like a wavenumber squared, making the longitudinal magnon effectively massive.
The critical longitudinal magnon, on the other hand, we expect to be massless. 
In this case, we choose $a_{\ell,d-2}$ to be marginal for
$2<d<4$, and $a_{\ell,2}$ for $d>4$. This implies a critical time scale
\be
z_{\rm c} = \begin{cases}
            d & \text{for $2<d<4$}\cr
            4 & \text{for $d>4$}\quad,
            \end{cases}
\label{eq:4.6}
\ee
as well as a critical exponent $\eta \equiv \eta_{\ell} = \eta_t$,
\be
\eta = \begin{cases}
       4-d & \text{for $2<d<4$}\cr
       0   & \text{for $d>4$}\quad.
       \end{cases}
\label{eq:4.7}
\ee
$t$ is by definition the only relevant operator (apart from the
temperature/frequency and the external magnetic field) at a physical critical
fixed point, and its scale dimension determines the correlation length
exponent $\nu = 1/[t]$. We thus have
\be
\nu = \begin{cases}
      1/(d-2) & \text{for $2<d<4$}\cr
      1/2     & \text{for $d>4$}\quad.
      \end{cases}
\label{eq:4.8}
\ee

We now turn to the scale dimension of $m$. From Eq.\ (\ref{eq:3.1c}), in
conjunction with Eqs.\ (\ref{eq:4.3a}) and (\ref{eq:4.4}), we have
\bse
\label{eqs:4.9}
\be
[m] = 2\quad.
\label{eq:4.9a}
\ee
However, in order to determine the physical order parameter exponents $\beta$
and $\delta$, we need to take into account
that the critical behavior of the magnetization is affected by dangerous
irrelevant variables. From the equation of state, Eq.\ (\ref{eq:2.9b}),
we see that $m(h=0) \propto v^{-2/(d-2)}$ for $2<d<6$, and
$m(h=0) \propto u^{-1/2}$ for $d>6$.
\footnote{We note that it is permissible at this point to use the generalized
 mean-field equation of state, whose validity we want to establish. The point
 is that the equation of state serves only to obtain the values of the
 exponents $\beta$ and $\delta$ from $[m]$, while the fact
 that $[m] = 2$ suffices to justisfy the scaling arguments used in I, which
 in turn establishes the validity of the equation of state up to logarithmic
 corrections. See also the last paragraph of this section.}
The effective scale dimension of $m$
is therefore $[m]_{\rm eff} = [m] + 2[v]/(d-2)$ for $2<d<6$, and
$[m]_{\rm eff} = [m] + [u]/2$ for $d>6$. But from the equation of state
we have $[u] = [t] - 2[m] = -2$ and $[v] = [t] - (d-2)[m]/2$. This yields
$[u] = -2$ for $d>6$, and $[v] = 0$ for $2<d<4$, and $[v] = 4-d$ for
$4<d<6$. Therefore,
\be
 [m]_{\rm eff} = \begin{cases}
                 2 & \text{for $2<d<4$}\cr
                 4/(d-2) & \text{for $4<d<6$}\cr
                 1 & \text{for $d>6$}\quad,
                 \end{cases}
\label{eq:4.9b}
\ee
which leads to
\be
\beta = \begin{cases}
        2/(d-2) & \text{for $2<d<6$}\cr
        1/2     & \text{for $d>6$}\quad.
        \end{cases}
\label{eq:4.9c}
\ee
\ese

Finally, the effective scale dimension of the external magnetic field $h$
is
\bse
\label{eqs:4.10}
\be
[h]_{\rm eff} = [t] + [m]_{\rm eff} = \begin{cases}
                d & \text{for $2<d<4$}\cr
                2d/(d-2) & \text{for $4<d<6$}\cr
                3 & \text{for $d>6$}\quad,
                \end{cases}
\label{4.10a}
\ee
which implies
\be
\delta = \begin{cases}
         d/2 & \text{for $2<d<6$}\cr
         3   & \text{for $d>6$}\quad.
         \end{cases}
\label{eq:4.10b}
\ee
\ese

An inspection of Eq.\ (\ref{eq:3.5}) shows that all corrections to the
Gaussian action are irrelevant with respect to the Gaussian fixed point, except
that the $c_2$ are marginal in the event that the factor $\sqrt{T}$ in
this coupling carries the diffusive time scale. It was shown in I and II
that this can indeed happen, and that this makes $c_2$ marginally relevant
with respect to the Gaussian fixed point. The actual critical fixed point
therefore contains the effects of $c_2$. This is a result of the existence
of two time scales in the problem. Furthermore, the terms of $O(q^4)$,
which are irrelevant by power counting, turn out to be effectively marginal
as well. This was also shown in I and II, and the logarithmic corrections
to scaling that result from these marginal operators were explicitly
determined.

There is no need to repeat the discussion of the logarithmic corrections to
scaling, since it
is now clear how this solution carries over to the present case. Above we have
shown explicitly that scaling works in the ferromagnetic phase, with the
magnetization having a scale dimension $[m] = 2$. This means in particular
that the results obtained in II from scaling arguments for the free energy
were correct. For instance, the exponent $\beta$ is given by the result
of the generalized Landau theory, Eq.\ (\ref{eq:2.10}), with logarithmic
corrections as given in Eq.\ (3.6e) of II.

\section{Discussion and Conclusion}
\label{sec:V}

We conclude with a summary of our results, and some additional remarks.

\subsection{Summary}
\label{subsec:V.A}

In summary, we have constructed an effective theory for the instability of
the ferromagnetic phase of a disordered itinerant Heisenberg ferromagnet at
the quantum critical point. We have shown that the presence of ferromagnetic
Goldstone modes, or spin waves, does not change the critical behavior compared
to the one obtained by supplementing results from the paramagnetic phase with
scaling arguments. We have also given a very simple generalized Landau theory
for this problem, which takes into accout the effects of soft fermionic modes
independent of the order parameter, and which yields the correct critical
behavior in all dimensions $d>2$ apart from logarithmic corrections to
power-law scaling. The results of the Landau theory are the same as those
originally obtained from a nonlocal order parameter theory in
Ref.\ \onlinecite{us_fm_dirty}.

\subsection{Hertz's fixed point}
\label{subsec:V.B}

We briefly discuss how Hertz's fixed point\cite{Hertz_76} relates to the
above discussion. Suppose one ignored the mode-mode coupling effects that
are represented by the coefficients $a_{\ell,d-2}$ and $a_{t,d-2}$ in the
Gaussian action, Eqs.\ (\ref{eqs:3.1}), and by $v$ in the equation of state,
Eq.\ (\ref{eq:2.9b}). Then one has, for all $d>2$,
\bse
\label{eqs:5.1}
\be
[t] = [m] = 2\quad,\quad [u] = -2\quad,
\label{eq:5.1a}
\ee
which leads to
\be
[m]_{\rm eff} = 1\quad,\quad [h]_{\rm eff} = 3\quad.
\label{eq:5.1b}
\ee
\ese
The dynamical critical exponent is then
\bse
\label{eqs:5.2}
\be
z_{\rm c} = 4\quad,
\label{eq:5.2a}
\ee
and all static exponents have mean-field values,
\be
\eta = 0\quad,\quad \nu = \beta = 1/2\quad,\quad \delta = 3\quad.
\label{eq:5.2b}
\ee
\ese
Of course, for $2<d<4$ this fixed point is unstable against the mode-mode
coupling effects as was discussed in detail in I. For $4<d<6$ it is actually
stable, and the only reason why Hertz's theory does not yield the correct
critical behavior is that it misses the leading dangerous irrelevant variable
for the magnetization, which is $v$ rather than $u$. For $d>6$ the exact
critical behavior is mean-field like.

\subsection{General remarks}
\label{subsec:V.C}

We finally come back to some of the points mentioned in the Introduction.
We have shown, by explicitly considering the ordered phase, that the scaling
arguments used in I and II to extract the critical behavior of the magnetization
were correct, and can be justified by a RG analysis. In particular, the
presence of magnetic Goldstone modes does not invalidate these arguments
since the transverse fermionic modes and the magnetic Goldstone modes
essentially just switch roles as one goes into the ferromagnetic phase.
We mention, however, that we have not addressed the question of the size
of the quantum scaling region for any particular observable. This question
can only be answered by an explicit crossover calculation for the
observable of interest, which follows the behavior of the observable through
the entire critical region at $T>0$.

Another point of interest is the relation between the theory developed here
and Ref.\ \onlinecite{us_fm_dirty}. Like Hertz's theory, the latter was a pure
Landau-Ginzburg-Wilson theory, i.e., a field theory formulated entirely in terms
of the order parameter field; the fermionic degrees of freedom had been
integrated out. This integrating out of soft modes resulted in nonlocal
vertices that made the theory unsuitable for explicit calculations, but the
critical behavior could be extracted by a combination of power counting and
scaling arguments. It is interesting to see that the current series of
papers has completely vindicated this treatment. In particular, the equation
of state derived in Ref.\ \onlinecite{us_fm_dirty} was Eq.\ (\ref{eq:2.9a})
expanded in powers of $m^2$, and the critical behavior determined in the
nonlocal theory was indeed exact except for the logarithmic corrections to
scaling in $2<d<4$ that were discussed in II. It is also interesting to
see that the much simpler generalized Landau theory of Sec.\ \ref{subsec:II.B},
which is based on a Gaussian approximation,
reproduces the results of Ref.\ \onlinecite{us_fm_dirty}, and thus is also
exact except for the logarithms. The reason is that, as we have seen, only
two classes of non-Gaussian terms ultimately contribute to the critical fixed
point, namely, the terms of $O(Mq^2)$ and those of $O(q^4)$, and both of
these lead only to logarithmic corrections to the Gaussian critical behavior.
Section \ref{sec:II} therefore provides a very simple way to obtain the
essentially exact critical behavior, but of course this is clear only
{\it a posteriori} once the RG analysis has been performed.

\acknowledgments
We gratefully acknowledge discussions with Ted Kirkpatrick and John Toner.
This work was supported by the NSF under grant No. DMR-01-32555 and by
the NSF IGERT fellowship program, grant No. DGE-0114419.

\appendix
\section{Derivation of the effective action}
\label{app:A}

In this appendix we give the exact effective action at the Gaussian level,
as it emerges from a derivation from the microscopic theory. A schematic
version of this result that is sufficient for power counting was given
in Eqs.\ (\ref{eqs:3.1}), and motivated by
general arguments in Sec.\ \ref{sec:III}. We start with the general
field theory, Eqs.\ (2.10) of I. The first step is to find a saddle-point
solution that is appropriate for a ferromagnetic phase. One then
separates soft and massive modes, and integrates out the latter in
tree approximation. The result is the desired effective field theory
for the soft modes.

\subsection{Stoner saddle point}
\label{app:A.1}

Equations (2.10) of I allow for a homogeneous saddle-point solution that
reproduces Stoner theory. This is the saddle point considered in
Ref.\ \onlinecite{us_fm_mit_I}, supplemented with a saddle-point value
for the order parameter field $M$,
\bse
\label{eqs:A.1}
\bea
^i_rQ_{12}({\bm x})|_{\rm sp} &=& \delta_{12}[\delta_{r0}\delta_{i0}\,Q^0_{n_1}
	+ \delta_{r3}\delta_{i3}\, Q^3_{n_1}] \quad,
\label{eq:A.1a}
\\
^i_r\tilde\Lambda_{12}({\bm x})|_{\rm sp} &=& \delta_{12}
	[-\delta_{r0}\delta_{i0}\, i\, \Sigma_{n_1}
	+ \delta_{r3}\delta_{i3}\, i\, \Delta_{n_1}] \quad,
\nonumber\\
\label{eq:A.1b}
\\
^iM_{1}({\bm x})|_{\rm sp} &=& \delta_{n_10}\delta_{i3}\,m/\sqrt{T} \quad.
\label{eq:A.1c}
\eea
\ese
Here we use the same notation as in I, with $1\equiv (n_1,\alpha_1)$, etc.,
comprising both frequency and replica indices.
By substituting Eqs.\ (\ref{eqs:A.1}) into the effective action ${\cal A}$
(Eqs.\ (2.10) in paper I),  and using the
saddle-point conditions $\delta{\cal A}/\delta Q=\delta{\cal A}/\delta
\tilde\Lambda = \delta{\cal A}/\delta M = 0$, one obtains the saddle-point
equations
\bse
\label{eqs:A.2}
\bea
m &=& 4i\left(\frac{\Gamma_t }{2}\right)^{1/2}T\sum_m Q^3_m e^{i \omega_m 0}
	\quad , \label{eq:A.2a} \\
Q^0_n &=& \frac{i}{ 2V}\sum_{\bm k}G^0_n({\bm k}) \quad , \label{eq:A.2b} \\
Q^3_n &=& \frac{i}{ 2V}\sum_{\bm k}G^3_n({\bm k}) \quad , \label{eq:A.2c} \\
\Sigma_n &=& \frac{-i}{ \pi N_{\rm F} \tau_{\rm el}}Q^0_n
	- 4iT\Gamma_s\sum_mQ^0_me^{i\omega_m0} \quad , \label{eq:A.2d} \\
\Delta_n &=& \frac{i}{ \pi N_{\rm F} \tau_{\rm el}}Q^3_n
	+ \Delta \quad,
\label{eq:A.2e}
\eea
with
\bea
\Delta = - (2\Gamma_t)^{1/2}m \quad .
\label{eq:A.2f}
\eea
\ese
Here
\bse
\label{eqs:A.3}
\bea
G^0_n({\bm k}) &=& \frac{1}{ 2}[{\cal G}^+_n({\bm k}) + {\cal G}^-_n({\bm k})]
	\quad,
\label{eq:A.3a} \\
G^3_n({\bm k}) &=& \frac{1}{ 2}[{\cal G}^+_n({\bm k}) - {\cal G}^-_n({\bm k})]
	\quad,
\label{eq:A.3b}
\eea
are Green functions given in terms of
\bea
{\cal G}^{\pm}_n({\bm k}) = \frac{1 }{ i\omega_n - \xi_{\bm k} \pm \Delta_n
	- \Sigma_n} \quad,
\label{eq:A.3c}
\eea
\ese
with $\xi_{\bm k} = {\bm k}^2/2m - \mu$.
$\Gamma_{s,t}$ are the interaction amplitudes proportional to $K_{s,t}$
that were defined in I. Upon substituting
Eqs.\ (\ref{eq:A.2a}) and (\ref{eq:A.2f}) into
(\ref{eq:A.2e}), one recovers the saddle-point equations of
Ref.\ \onlinecite{us_fm_mit_I}, and hence Stoner theory.

It is useful to define various transport and thermodynamic quantities
in terms of these Green functions. We will need
\bse
\label{eqs:A.4}
\bea
\sigma_0^{\pm} &=& \frac{1}{\pi m V} \lim_{n,m\rightarrow 0}\sum_{\bm k}
                 \left[\frac{1}{2}\left({\cal G}_n^{\pm}({\bm k}) +
                 {\cal G}_m^{\pm}({\bm k})\right) \right.
\nonumber\\
&& + \left. \frac{1}{dm}\,{\bm k}^2\,
                 {\cal G}_n^{\pm}({\bm k})\,{\cal G}_m^{\pm}({\bm k})\right]
                 \quad,
\label{eq:A.4a}
\eea
and
\bea
{\tilde\sigma}_0^{\pm} &=& \frac{1}{\pi m V} \lim_{n,m\rightarrow 0}\sum_{\bm k}
                 \left[\frac{1}{2}\left({\cal G}_n^{\pm}({\bm k}) +
                 {\cal G}_m^{\mp}({\bm k})\right) \right.
\nonumber\\
&&+ \left.  \frac{1}{dm}\,{\bm k}^2\,
                 {\cal G}_n^{\pm}({\bm k})\,{\cal G}_m^{\mp}({\bm k})\right]
                 \quad,
\label{eq:A.4b}
\eea
as well as
\be
N_{\rm F}^{\pm} = \frac{i}{\pi V}\lim_{n\rightarrow 0+}\sum_{\bm k}
                  {\cal G}_n^{\pm}({\bm k})\quad.
\label{eq:A.4c}
\ee
\ese
These quantities represent the Born approximation for various conductivities
and densities of states in the split-band system of Stoner theory. They are
generalizations of the analogous quanties defined in
Ref.\ \onlinecite{us_fermions_I}. For a physical interpretation of these
quantities, see Ref.\ \onlinecite{sls_thesis}.

\subsection{Gaussion soft-mode theory}
\label{app:A.2}

The separation of soft and massive modes works in analogy to I, although
the procedure is more cumbersome in the presence of a nonzero magnetization.
The massive modes are integrated out in tree approximation to arrive at
an effective soft-mode action,
and the soft fermion modes are expanded in powers
of $q$ (cf. Eq.\ (\ref{eq:2.5d})), again in analogy to I.
The complete procedure and result can be found in Ref. \onlinecite{sls_thesis}.
Here we list the Gaussian (i.e., bilinear in ${\bm M}$ and
$q$) contribution to the effective action. The higher order terms are
obtained analogously.

The Gaussian action has the general form
\be
{\cal A}^{\rm G}[q,M] = {\cal A}_{\rm NL\sigma M}[q] + {\cal A}_{\rm LGW}[M]
	+{\cal A}_c[q,M]\quad.
	\label{eq:A.5}
\ee
We expand the $q$ matrices in a spin-quaternion basis, see
Eq.\ (\ref{eq:2.5b}). The fermionic part of the action, which is the Gaussian
part of a generalized nonlinear sigma model, is given by
\begin{widetext}
\bea
{\cal A}_{\rm NL\sigma M}[q] &=& -4\sum_{\alpha\beta}\sum_{n_1,n_2,n_3,n_4}
	\sum_{\bm k}\sum_{r,s=0,3}\sum_{i,j=0}^3
	\biggl[
	\delta_{n_1n_3}\delta_{n_2n_4}\Bigl[
	\kappa^0_{ij,rs}\,(\delta_{i0} + \delta_{i3})
	\left({\bm k}^2/G^{\rm s} + H^{\rm s}\Omega_{n_1-n_2}\right)
	\nonumber \\
&&
	+ \kappa^0_{ij,rs}\,(\delta_{i1} + \delta_{i2})
              \left({\bm k}^2/{\tilde G}^{\rm s}
                        + H^{\rm s}\Omega_{n_1-n_2}\right)
        + \kappa_{ij,rs}^{\ell}
	\left({\bm k}^2/G^{\rm a} + H^{\rm a}\Omega_{n_1-n_2}\right)
	+ \kappa_{ij,rs}^{t}
	 \left({\bm k}^2/\tilde{G}^{a} +2iH^{\rm s}\Delta\right)\Bigr]
	\nonumber \\
&& \hskip -50pt
	- \delta_{n_1-n_2,n_3-n_4}\delta_{\alpha\beta}\,
	 T\Gamma_s \pi^2
	 \Bigl[ \kappa_{ij,rs}^0 \left[
	(N_{\rm F}^{\rm s})^2\ \delta_{i0}
	+ (N_{\rm F}^{\rm a})^2\ \delta_{i3} \right]
	+\ \kappa_{ij,rs}^{\ell}N_{\rm F}^{\rm s} N_{\rm F}^{\rm a} \Bigr]
	\biggr]
	 \ ^i_rq_{n_1n_2}^{\alpha\beta}({\bm k})\
	^j_sq_{n_3n_4}^{\alpha\beta}(-{\bm k}) \quad .
\label{eq:A.6}
\eea
Here the $\kappa$ represent traces in spin-quaternion space,
\bse
\label{eqs:A.7}
\bea
\kappa_{ij,rs}^0 &=& \frac{1}{4} \tr (\tau_r\tau_s^{\dagger})
	\tr (s_is_j^{\dagger}) \quad,\quad
\kappa_{ij,rs}^{\ell} =  \frac{1}{8} \tr (\tau_3\tau_r\tau_s^{\dagger})
	\left[ \tr (s_3s_is_j^{\dagger}) +
	\tr (s_is_3s_j^{\dagger}) \right] \quad,
\label{eq:A.7a}\\
\kappa_{ij,rs}^{t}&=& \frac{1}{8} \tr (\tau_3\tau_r\tau_s^{\dagger})
	\left[ \tr (s_3s_is_j^{\dagger}) -
	\tr (s_is_3s_j^{\dagger}) \right] \quad,\quad
\kappa_{ij,rs}^3  =  -\frac{1}{4} \tr (\tau_r\tau_s^{\dagger})
	\tr (s_3s_is_3s_j^{\dagger}) \quad.
\label{eq:A.7b}
\eea
\ese
The coupling constants $G$ and $H$ in Eq.\ (\ref{eq:A.6}), as well as
the densities of states $N_{\rm F}$, are magnetization dependent
generalizations of the analogous quantities in I. They are given by
\bse
\label{eqs:A.8}
\bea
G^{\rm s} &=& 16/\pi (\sigma_0^+ + \sigma_0^-)\quad,\quad
G^{\rm a}  =  16/\pi (\sigma_0^+ - \sigma_0^-)\quad,\quad
{\tilde G}^{\rm s}  =  16/\pi ({\tilde\sigma}_0^+ + \tilde{\sigma}_0^-)\quad,
\quad
{\tilde G}^{\rm a}  =  16/\pi ({\tilde\sigma}_0^+ - \tilde{\sigma}_0^-)\ ,
\label{eq:A.8a} \\
N_{\rm F}^{\rm s} &=& (N_{\rm F}^+ + N_{\rm F}^-)/2\quad,\quad
N_{\rm F}^{\rm a}  =  (N_{\rm F}^+ - N_{\rm F}^-)/2\quad,\quad
\label{eq:A.8b}\\
H^{\rm s,a} &=& \pi\,N_{\rm F}^{\rm s,a}/4\quad.
\label{eq:A.8c}
\eea
\ese
As can be seen by a
direct comparison, the schematic action given by Eq.\ (\ref{eqs:3.1})
reflects all qualitative features of the complete result. Notice that
the mass, ${H}\Delta$, in the transverse vertex
comes from the coupling between $M$ and $q$, cf. Eq.\ (\ref{eq:2.4}),
but has been included in ${\cal A}_{{\rm NL}\sigma{\rm M}}$ since it
contains only fermionic fluctuations.

The LGW part of the action that results from integrating out the massive modes
is
\bea
{\cal A}_{\rm LGW}[M] &=& -
	\sum_{\alpha}\sum_{n\ge 0}(2- \delta_{n0})
	\sum_{\bm k}\sum_{i,j=1}^3
	\Bigg\{ \delta_{ij}
	+\ (\delta_{i1}\delta_{j1}+\delta_{i2}\delta_{j2}
	-i\kappa_{12,ij}^t
	)\Gamma_tT{\sum_m}'
	\left( {\cal E}^{+}_{m+n,m} + (-)^{i+j}
	{\cal E}^{-}_{m+n,m} \right)
\nonumber\\
	&&\hskip -30pt +\ \delta_{i3}\delta_{j3}\Gamma_tT{\sum_m}'
	\left( {\cal D}^{+}_{m+n,m}
	+ {\cal D}^{-}_{m+n,m} \right)
	+\ \delta_{i3}\delta_{j3}\frac{\Gamma_t\Gamma_s
	\left[ T\sum_m'
	\left({\cal D}^{+}_{m+n,m} - {\cal D}^{-}_{m+n,m}\right)\right] ^2
	}{ 1-\Gamma_sT\sum_m'
	\left( {\cal D}^{+}_{m+n,m} + {\cal D}^{-}_{m+n,m} \right) }
	\Bigg\}
	\ ^iM^{\alpha}_n({\bm k})\ ^jM^{\alpha}_{-n}(-{\bm k})
	\ .
	\nonumber\\
&&\label{eq:A.9}
\eea
\end{widetext}
Here, we have adopted the notation of Ref. \onlinecite{us_fm_mit_I}; ${\cal D}$
and ${\cal E}$ correspond to modified ``diffusons" in the longitudinal and
transverse spin channels, respectively.  They are given by
\bse
\bea
{\cal D}^{\pm}_{nm}({\bm k}) &=& \phi^{\pm}_{nm}({\bm k})/
	[1-\phi^{\pm}_{nm}({\bm k})/2\pi N_{\rm F}\tau]\ ,\label{eq:A.10a}\\
{\cal E}^{\pm}_{nm}({\bm k}) &=& \eta^{\pm}_{nm}({\bm k})/
	[1-\eta^{\pm}_{nm}({\bm k})/2\pi N_{\rm F}\tau] \ ,\label{eq:A.10b}
\eea
with
\bea
\phi^{\pm}_{nm} &=& (\varphi^{00}_{nm} + \varphi^{33}_{nm}) \pm
		    (\varphi^{03}_{nm} + \varphi^{30}_{nm})
		    \quad, \label{eq:A.10c}\\
\eta^{\pm}_{nm} &=& (\varphi^{00}_{nm} - \varphi^{33}_{nm}) \pm
		    (\varphi^{03}_{nm} - \varphi^{30}_{nm})
		    \quad, \label{eq:A.10d}
\eea
and
\be
\varphi^{uv}_{nm}=\frac{1}{V}\sum_{\bm p}G_n^{u}({\bm p})\,
	G_m^{v}({\bm p}+{\bm k}) \quad;\quad (u,v=0,3) \quad,
	\label{eq:A.10e}
\ee
\ese
with $G^{0,3}$ given by Eqs.\ (\ref{eqs:A.3}).
We have also introduced the notation
\be
T{\sum_m}' [\ \ldots\ ] \equiv
	T\sum_m \Theta (m(m+n))\ [\ \ldots\ ]\quad.
\label{eq:A.11}
\ee

The LGW vertex simplifies in the limit of long wavelengths and small
frequencies. By using\cite{us_fm_mit_I}
\bse
\label{eqs:A.12}
\be
T\sum_m {\cal D}_{mm}^{\pm}({\bm k}=0) = -N_{\rm F}^{\pm}\quad,
\label{eq:A.12a}
\ee
and
\be
T\sum_m {\cal E}_{mm}^{\pm}({\bm k}=0) = -1/2\Gamma_t\quad,
\label{eq:A.12b}
\ee
\ese
one recovers the functional form given in Eqs.\ (\ref{eqs:3.1}), except
for the nonanalytic terms that emerge only at one-loop order and were
added in Sec.\ \ref{sec:III}.

Finally, the coupling between $M$ and $q$ to bilinear order is
\bse
\label{eqs:A.13}
\bea
{\cal A}_c[M,q] &=&
        \sum_{12}\sum_{\bm k}\sum_{r,s=0,3}\sum_{i=1}^3\sum_{j=0}^3
        4\pi\sqrt{2T\Gamma_t} \nonumber \\
&& \hskip -50pt
        \times\ \Big[ \kappa_{ij,rs}^0(N_{\rm F})^2 +
        \kappa_{ij,rs}^{\ell}(N_{\rm F}^a)^2 \Big]
        \ ^i_rb_{12}({\bm k})\ ^j_sq_{12}(-{\bm k}) \quad.
        \nonumber\\
&&
        \label{eq:A.13a}
\eea
Here, $b$ is a field with components
\bea
^i_rb_{12}({\bm k})&=&\delta_{\alpha_1\alpha_2}(-)^{r/2}\sum_n
        \delta_{n,n_1-n_2}\big[\, ^i M^{\alpha_1}_n({\bm k})
        \nonumber\\
&&\hskip 20pt
        +(-)^{r+1}\, ^iM^{\alpha_1}_{-n}({\bm k})\quad.
\label{eq:A.13b}
\eea
\ese
Again, this has the same functional form as the schematic representation
in Eqs.\ (\ref{eqs:3.1}).

\subsection{Gaussian propagators}
\label{subsec:A.3}

The Gaussian propagators can be obtained by inverting the quadratic form
given by the Gaussian action in the previous subsection. We do not list
the complete propagators here, which can be found in
Ref.\ \onlinecite{sls_thesis}, but give only the hydrodynamic parts of
the diffusons, which are needed in the calculation. For $nm<0$ one finds
\bse
\label{eqs:A.14}
\be
{\cal D}^{\pm}_{nm}({\bm k}) = \frac{2\pi N_{\rm F}^{\pm} }{ D^{\pm}{\bm k}^2 +
	|\Omega_{n-m}|} \quad .
\label{eq:A.14a}
\ee
Here, $D^{\pm}=1/G^{\pm}H^{\pm}$ with
\bea
1/G^{\pm} &=& 1/G^{\rm s} \pm 1/G^{\rm a}\quad,
\label{eq:A.14b}\\
H^{\pm} &=& H^{\rm s} \pm H^{\rm a}\quad,
\label{eq:A.14c}
\eea
\ese
can be interpreted as the Boltzmann diffusivities in the upper and lower
Stoner band, respectively.

Similarly, one finds for $n>0$ and  $m<0$,
\bse
\label{eqs:A.15}
\be
{\cal E}^{\pm}_{nm}({\bm k}) = \frac{2\pi N_{\rm F}}{
        [\Omega_{n-m} + \tilde{D}^{\pm}{\bm k}^2]
	\frac{1}{ 1\pm 2i\Delta\tau} \pm 2i\Delta} \quad.
\label{eq:A.15a}
\ee
Here, $\tilde{D}^{\pm}=D/(1\pm 2i\Delta\tau)=1/\tilde{G}^{\pm}H$,
with
\be
1/{\tilde G}^{\pm} = 1/{\tilde G}^{\rm s} \pm 1/{\tilde G}^{\rm a}\quad.
\label{eq:A.15b}
\ee
Notice that Eq.\ (\ref{eq:A.15a}) holds only for $n>0$ and $m<0$.
The result for $n<0$, $m>0$ is given by the relation
\footnote{This distinction between the cases $n>0$, $m<0$ and $n<0$, $m>0$
 was not made in Ref. \onlinecite{us_fm_mit_I}.}
\be
{\cal E}_{nm}^{\pm}({\bm k}) = {\cal E}_{mn}^{\mp}({\bm k})\quad.
\label{eq:A.15c}
\ee
\ese

By means of these expressions, one readily finds that the propagators have
functional forms as given in Sec.\ \ref{sec:III}. Again, the nonanalytic
dependences on ${\bm k}$ and $\Delta$ are not included in the Gaussian
theory, and have been added in Sec.\ \ref{sec:III}.

%\bibliography{paper_III}

\end{document}